# Spectator-transition crosstalk in a spin-3/2 silicon-vacancy qudit in silicon carbide revealed by broadband Ramsey interferometry

Jun-Jae Choi,[1,†] Seung-Jae Hwang,[1,†] Seoyoung Paik,[1] Juhwan Kim,[2] Jawad Ul-Hassan,[3] Nguyen Tien Son,[3] Hiroshi Abe,[4] Takeshi Ohshima,[4,5] Jaekwon Suk,[6] Hyeon-Ho Jeong,[2] Dong-Hee Kim,[1] Sang-Yun Lee[1,*]

*[1] Department of Physics and Photon Science, Gwangju Institute of Science and Technology, Cheomdan-gwagiro 123, Gwangju, Republic of Korea*
*[2] Department of Electrical Engineering & Computer Science, Gwangju Institute of Science and Technology, Cheomdan-gwagiro 123, Gwangju, Republic of Korea*
*[3] Department of Physics, Chemistry and Biology, Linköping University, SE-581 83 Linköping, Sweden*
*[4]National Institutes for Quantum Science and Technology (QST), Takasaki, Gunma 370-1292, Japan*
*[5]Department of Materials Science, Tohoku University, Sendai, Miyagi 980-8579, Japan*
*[6] Korea Multi-purpose Accelerator Complex, Korea Atomic Energy Research Institute, Gyeongju, 38180, Republic of Korea*

Color-center spins in 4H-SiC offer a rare combination of wafer-scale materials maturity with long spin coherence and chip-level photonics, making them promising building blocks for scalable quantum technologies. In particular, the silicon vacancy hosts an $S = 3/2$ ground state, a native qudit that enables compact encodings and subspace-selective control but also introduces spectator transitions: short, detuned pulses can coherently drive nonaddressed level pairs and create crosstalk. Here we use broadband Ramsey interferometry to reveal and quantify such spectator-transition crosstalk. Experimentally, the Ramsey Fourier spectra display multiple lines beyond the addressed single-quantum transition. Analytically, we map each line to a pairwise energy difference between qudit levels of the rotating-frame Hamiltonian and assign its weight via compact amplitudes set by the prepared state and the microwave pulse parameters, predicting a deterministic six-branch structure. Numerical time-domain propagation with the experimental sampling reproduces the detuning map, and the measured peak positions coincide with the analytic branch lines without frequency fitting. Together these results provide a practical, spectator-aware framework for multilevel control in the silicon-vacancy qudit. The approach offers clear guidance to suppress crosstalk or, conversely, to exploit spectator lines, e.g., as additional constraints for in situ pulse calibration and for phase-sensitive quantum state and process estimation.

† These authors contributed equally
* sangyunlee@gist.ac.kr

## I. INTRODUCTION.

Color centers in 4H-SiC have rapidly matured into a semiconductor platform for quantum technologies. Leading contenders such as the silicon vacancy and the divacancy combine long-lived, optically addressable spins with high-quality nanophotonics, bringing high-fidelity spin–photon interfaces within practical reach and, by extension, quantum-repeater and network architectures [1,2]. Recent milestones highlight device-level control of essential quantum properties of these centers: standard optoelectronic structures, e.g., Schottky and PIN junctions, can tune the local charge environment and Stark-shift single defects, narrowing zero-phonon lines (ZPLs) toward the lifetime limit required for generating indistinguishable photons in Hong–Ou–Mandel Bell-state measurements, as well as enhancing spin coherence [3-5]. In parallel, nanophotonic waveguides and cavities preserve lifetime limited optical transitions and robust spin control, including coherent manipulation of nearby nuclear registers [2]. Thin-film SiC-on-insulator studies have chiefly showcased photonic integration and on-chip frequency conversion [6]. While these advances have been driven by various color centers, the negatively charged silicon vacancy stands out because its optical transition dipole has a single orientation aligned with the crystal c-axis [7,8], which simplifies deterministic coupling to nanophotonic modes, and its spin quantization axis is also along the c-axis for all sites [9,10], which supports scalable multispin registers and higher multiqubit gate fidelities. Also, its optical lines remain close to the lifetime limit after nanofabrication and up to 20 K [2,11].

A distinctive feature of the negatively charged silicon vacancy in 4H-SiC is its spin-3/2 ground state [12,13], which realizes a natural four-level qudit, i.e., a ququart [14]. Such a multilevel structure can support higher information density per physical carrier [15] and reduced circuit depth in some architectures [16,17], hardware-efficient quantum error correction [18], and multitransition sensing in time-domain interferometry [19]. At the same time, closely spaced transitions make the system susceptible to off-resonant driving: bandwidth-broad microwave (MW) pulses can unintentionally excite spectator transitions, nonaddressed transitions that acquire coherence under off-resonant coupling [20]. This produces coherent leakage, i.e., drive-induced population transfer outside the intended subspace [21], and crosstalk, i.e., unintended excitation or correlated errors mediated by residual couplings or control spillover [22]. These considerations are not unique to any single platform since most physical platforms used for quantum information, including superconducting circuits [23], trapped ions and neutral atoms [24,25], and defect spins [26], are inherently multilevel even when treated as qubits. Reducing such systems to two levels simplifies control and modeling but can hide off-resonant excitation and state mixing that limit gate and readout fidelities [27], bias tomographic reconstruction [28], and complicate error characterization by introducing correlated, crosstalk-mediated errors [22]. Such effects grow with spin quantum number and with the use of spectrally broad, time-efficient pulses. These benefits and challenges motivate explicit modeling and experimental diagnosis of multilevel pathways of the silicon vacancy in 4H-SiC rather than qubit-only approximations.

Optically detected magnetic resonance (ODMR) provides the measurement backbone: continuous-wave ODMR maps transition frequencies and the ZFS, while pulsed ODMR enables time-domain protocols such as Rabi, Ramsey, and Hahn-echo. Among these, Ramsey interferometry is phase-sensitive to detuning and, with short hard $\frac{\pi}{2}$ pulses, its spectral bandwidth overlaps neighboring transitions, making it a sensitive probe of multilevel subspaces. In practice, Ramsey-type measurements are a workhorse for phase-sensitive state characterization, including estimation of detunings and coherences, and are routinely embedded in quantum state/process tomography (QST/QPT) workflows across multiple platforms [29-32]. In this work, we implement broadband Ramsey spectroscopy to probe multilevel dynamics in the 4H-SiC silicon-vacancy ququart, resolving crosstalk and coherent spectator excitation among the sublevels. In 4H-SiC, the two crystallographic variants of the negatively charged silicon vacancy, $V1$ (h-site) and $V2$ (k-site), exhibit different ground-state ZFS: $2D_{gs}/2\pi \approx 4.5$ MHz and 70 MHz for $V1$ and $V2$, respectively [12,13]. We therefore use the $V1$ center as a testbed whose small ZFS brings neighboring transitions within the bandwidth of short $\pi/2$ pulses, enhancing Ramsey's sensitivity to off-resonant pathways without large frequency excursions. Below we refer to such off-resonant, nonaddressed sublevel pairs that nevertheless acquire coherence during the sequence as spectator transitions [20]. Using this choice, we show, via simulations supported by analytical modeling, that a resonant drive seeds coherent amplitudes in spectator transitions; during free precession the full four-level density matrix accumulates multiple relative phases across pairwise coherences, and the final $\pi/2$ pulse mixes a subset of these coherences into measurable population modulations, yielding detuning-dispersive, multifrequency Ramsey spectra. We map how detuning and drive strength redistribute spectral weight among the $m_S = \pm 3/2, \pm 1/2$ components and provide an interpretive framework for multifrequency Ramsey spectra in spin-3/2 systems, linking the six detuning-dependent frequency components resolved in our spectra to specific pairwise coherences and their phase evolution. Looking ahead, a phase-resolved, multifrequency view of Ramsey signals may help future efforts to better understand and exploit qudits for example by informing gate design and error budgets, supporting sensing strategies that leverage sublevel-dependent shifts, or, where appropriate, augmenting measurement sets in state or process estimation.

## II. METHODS

### A. Sample and experimental setup

The sample used in this study is a 28 μm thick 4H-$^{28}$Si$^{12}$C silicon carbide layer grown by chemical vapor deposition on an n-type a-plane 4H-SiC wafer. The isotope purity of the epitaxially grown layer is ~99.85% for $^{28}$Si and ~99.98% for $^{12}$C as determined from secondary ion mass spectrometry for a representative wafer from the same growth series. The background doping concentration of the N shallow donor is in the range of ~$3\times10^{15}$ cm$^{-3}$. Prior to the silicon vacancy creation, the layer was annealed at 1150 °C for one hour to reduce surface defects. The isolated silicon vacancies were then introduced by electron irradiation. The sample was irradiated with 2 MeV electrons at a fluence of $1\times10^{12}$ cm$^{-2}$ with a flux of $4.8\times10^{10}$ cm$^{-2}$ s$^{-1}$ at room temperature yielding a low density of well-isolated silicon vacancies, approximately one V1 center per 1 μm$^2$ (estimated from confocal fluorescence scans). After irradiation, the sample was annealed at 400 °C for 30 min to remove some interstitial-related defects. In addition, 200 keV proton irradiation with a fluence of $1\times10^{15}$ cm$^{-2}$ was performed through a 60 μm thick aluminum mask to create an ensemble of silicon vacancies near one edge of the sample; this ensemble was used as a bright ensemble reference for the magnetic field alignment [9,33].

The experiments were performed in a cryogenic confocal microscope. A closed-cycle cryostat (Montana Instruments s200) was used to cool the sample to ~4-5 K. Inside the cryostat, a three-axis low-temperature nanopositioning stage was employed to scan the sample. We used a Zeiss EC Epiplan-Neofluar 100×, NA 0.75 air objective, with the excitation beam incident horizontally onto the sample. A 785 nm laser (Thorlabs L785H1) was used for depolarization of the ground state spin-sublevels and an 861.4 nm external-cavity diode laser (TOPTICA DL pro) for resonant excitation of the ZPL. The resonant laser frequency was stabilized using a wavemeter (HighFinesse WS-8-10) referenced to a frequency-stabilized He-Ne laser (632.991405 nm, ±5 MHz). Each laser passed through an acousto-optic modulator to enable pulsed operation for time-resolved experiments, e.g. Ramsey sequences, and the two beams were combined on a dichroic mirror (Thorlabs DMLP805). In the confocal microscope, the excitation and collection paths were separated by a wedged beam splitter (Thorlabs BSF2550). The collected fluorescence was filtered with a 875 nm long-pass filter (Thorlabs FELH0875) so that only photons from the red-shifted phonon sideband (PSB) were detected. Single photons were recorded using a superconducting nanowire single-photon detector (Single Quantum Eos R12). MW control was provided by a vector signal generator (Rohde & Schwarz SMBV100A). An FPGA-based control unit (National Instruments) was used to generate timing triggers for the lasers and MW pulses and to count photon events. A static magnetic field was applied by an external permanent magnet positioned outside the cryostat.

### B. Spin Hamiltonian

To model the coherent dynamics of the $V1$ center in 4H-SiC, whose ground state is a spin-$3/2$ quartet [12,13], we adopt a minimal Hamiltonian that captures the dominant energy terms and the interaction with a resonant MW drive. The spin-3/2 manifold comprises four Zeeman sublevels $(m_s = \pm 3/2, \pm 1/2)$. This enlarged Hilbert space enables rich multilevel dynamics but also admits weak off-resonant excitation of nonaddressed transitions, which can manifest as spectral crosstalk. We choose the crystal c-axis as the quantization axis $(z \parallel c)$. In the laboratory frame the ground state spin Hamiltonian is

$$\hat{H}_{\text{lab}} = \omega_0 \hat{S}_z + D_{gs}\left(\hat{S}_z^2 - \frac{5}{4}\hat{I}\right) + 2\omega_1 \cos(\omega t)\hat{S}_x, \tag{1}$$

where $\omega_0 = g\mu_B B_0/\hbar$ is the Larmor frequency for a static magnetic field $B_0$, $\omega_1 = g\mu_B B_1/\hbar$ is the rotating-frame microwave drive amplitude associated with the co-rotating field component $B_1$ such that the linearly oscillating laboratory-frame field is $B_x(t) = 2B_1\cos(\omega t)$. The constant $-5D_{gs}/4$ shifts all spin sublevels uniformly and does not affect transition frequencies or observable dynamics apart from an overall global phase.

Moving to the frame rotating at the drive frequency $\omega$ and applying the rotating-wave approximation (RWA) [34,35] yields

$$\hat{H}_{\text{rot}} = (\omega_0 - \omega)\hat{S}_z + D_{gs}\left(\hat{S}_z^2 - \frac{5}{4}\hat{I}\right) + \omega_1 \hat{S}_x. \tag{2}$$

Throughout, we define the detuning by $\delta \equiv \omega - (\omega_0 + 2D_{gs})$ so that $\delta > 0$ corresponds to a drive set above the $|+3/2\rangle \leftrightarrow |+1/2\rangle$ resonance frequency $(\omega_0 + 2D_{gs})$. When the $B_1$ drive is off during Ramsey free precession, i.e., $\omega_1 = 0$, $\hat{H}_{\mathrm{rot}}$ is diagonal in the $\{|+3/2\rangle, |+1/2\rangle, |-1/2\rangle, |-3/2\rangle\}$ basis with eigenvalues $E_{-3/2} = 3\delta/2 + 4D_{gs}$, $E_{-1/2} = \delta/2$, $E_{+1/2} = -\delta/2 - 2D_{gs}$, and $E_{+3/2} = -3\delta/2 - 2D_{gs}$. These level energies directly generate the detuning-dispersive frequency branches observed in the Fourier spectra via pairwise differences $\Omega_{mn} = |E_m - E_n|$ (detailed mapping appears in Sec. III D). The explicit $4 \times 4$ matrix forms can be found in Appendix A.

### C. Ramsey pulse sequence

We employ a standard Ramsey sequence, a widely used interferometric protocol in quantum information processing and quantum metrology, consisting of a $\pi/2$ pulse, a free-evolution interval $\tau$, and a second $\pi/2$ pulse with a fixed relative phase $\phi = 0$ with respect to the first pulse [35-37]. The drive is applied at frequency $\omega$ with detuning $\delta$. When optical spin polarization is implemented, the V1 center is initialized into a statistically equal mixture between $|+1/2\rangle$ and $|-1/2\rangle$ ground-state sublevels [13]. Although the selective initialization protocol has been well established both theoretically [38] and experimentally [12], we use the equal mixture of the $|\pm1/2\rangle$ sublevels of the $V1$ center as the initial state since we aim to observe various features arising from the crosstalk between sublevels. However, for the analytical expressions, we assume the initialization into $|+1/2\rangle$ because it is easier to handle and sufficient to find the analytical form for phase evolution as a result of the Ramsey sequence. The overall unitary is

$$\hat{U}_{Ramsey}(\tau) = \hat{R}_x\left(\frac{\pi}{2}\right) e^{-i\hat{H}_{rot}\tau} \hat{R}_x\left(\frac{\pi}{2}\right), \tag{3}$$

where $H_{rot}$ is defined as in Sec. II B, which becomes diagonal during free evolution $(\omega_1 = 0)$ time $\tau$. Then, the pairwise differences in the eigenvalues of the quartet sublevels generate the detuning-dispersive components, $\Omega_{mn}$, observed in the Fourier spectra. By applying the hard-pulse approximation, we can find the explicit form of $\hat{R}_x(\pi/2)$ (see Appendix B) which allows us to find the final state $|\psi_f\rangle$, explicitly [34]. Then, the final state projects onto $\hat{O}_2 = |-3/2\rangle\langle-3/2| + |+3/2\rangle\langle+3/2|$ since we use $O_2$ optical resonance for the optical spin population readout (see Fig. 1(a)) [38]. Then, the observable Ramsey signal $S(\tau, \delta) = \langle O_2 \rangle$ can be written as,

$$S(\tau) = \langle\psi_f|\hat{O}_2|\psi_f\rangle = C_0 + \sum_{m<n} 2\,|X_{mn}|\cos[\Omega_{mn}\tau - \arg X_{mn}], \tag{4}$$

where $C_0$ and $X_{mn}$ are constants determined by the initial state and the matrix elements of $\hat{R}_x(\pi/2)$. The exact definitions for $X_{mn}$ and explicit 4×4 matrix forms of $\hat{R}_x(\pi/2)$ and the detailed derivation are provided in Appendix B. The final solution, Eq. (4), leads to finding six values for $\Omega_{mn}$,

$$\{\Omega\} = \{\delta, \delta + 2D_{gs}, \delta + 4D_{gs}, 2\delta + 2D_{gs}, 2\delta + 6D_{gs}, 3\delta + 6D_{gs}\}, \tag{5}$$

which can also be found in the experimental data as will be explained below. In practice we sweep $\delta$ at fixed $B_0$ and $\tau$, compute the fast Fourier transform (FFT) of $S(\tau)$ to locate $\{\Omega_{mn}\}$, and compare the extracted peaks with the linear branches predicted above.

## III. RESULTS

### A. Optical resonances and coherent control of the $S = 3/2$ ground manifold

To support our crosstalk analysis of the $V1$ center's $S = 3/2$ ground-state manifold in 4H-SiC, we first establish the optical resonances used for initialization/readout and verify coherent spin control. These measurements set the model parameters for the simulations that follow and determine the experimental settings required for high-visibility Ramsey measurements. We operate at $T \approx 5$ K so that photoluminescence-excitation (PLE) spectra are sufficiently narrow to resolve the two spin-selective ZPL resonances cleanly, enabling clean resolution of two spin-preserving ($\Delta m_S = 0$) and spin-selective ZPL resonances [12,39].

We locate these lines by scanning the laser across the ZPL while monitoring PSB emission. As shown in Fig. 1(b), the PLE spectrum exhibits two narrow resonances separated by $\approx$ 1 GHz, consistent with the excited-state zero-field splitting $D_{es}$ [11,12,39,40]. The PLE spectrum shown in Fig. 1(b), measured under weak resonant excitation at 90 nW in front of the objective lens, exhibits the narrowest observed linewidth for this emitter, 170 $\pm$ 5 MHz (FWHM), allowing clear separation of the two resonant peaks and robust spin-selective addressing [12]. In our notation, $O_1$ denotes the pair $| g, m_S = \pm 1/2\rangle \leftrightarrow | e, m_S = \pm 1/2\rangle$ and $O_2$ denotes $| g, m_S = \pm 3/2\rangle \leftrightarrow | e, m_S = \pm 3/2\rangle$ (see Fig. 1(a)) [40]. These observations are consistent with a stable charge state under our conditions and minimal inhomogeneous broadening from nearby defects [11,41].

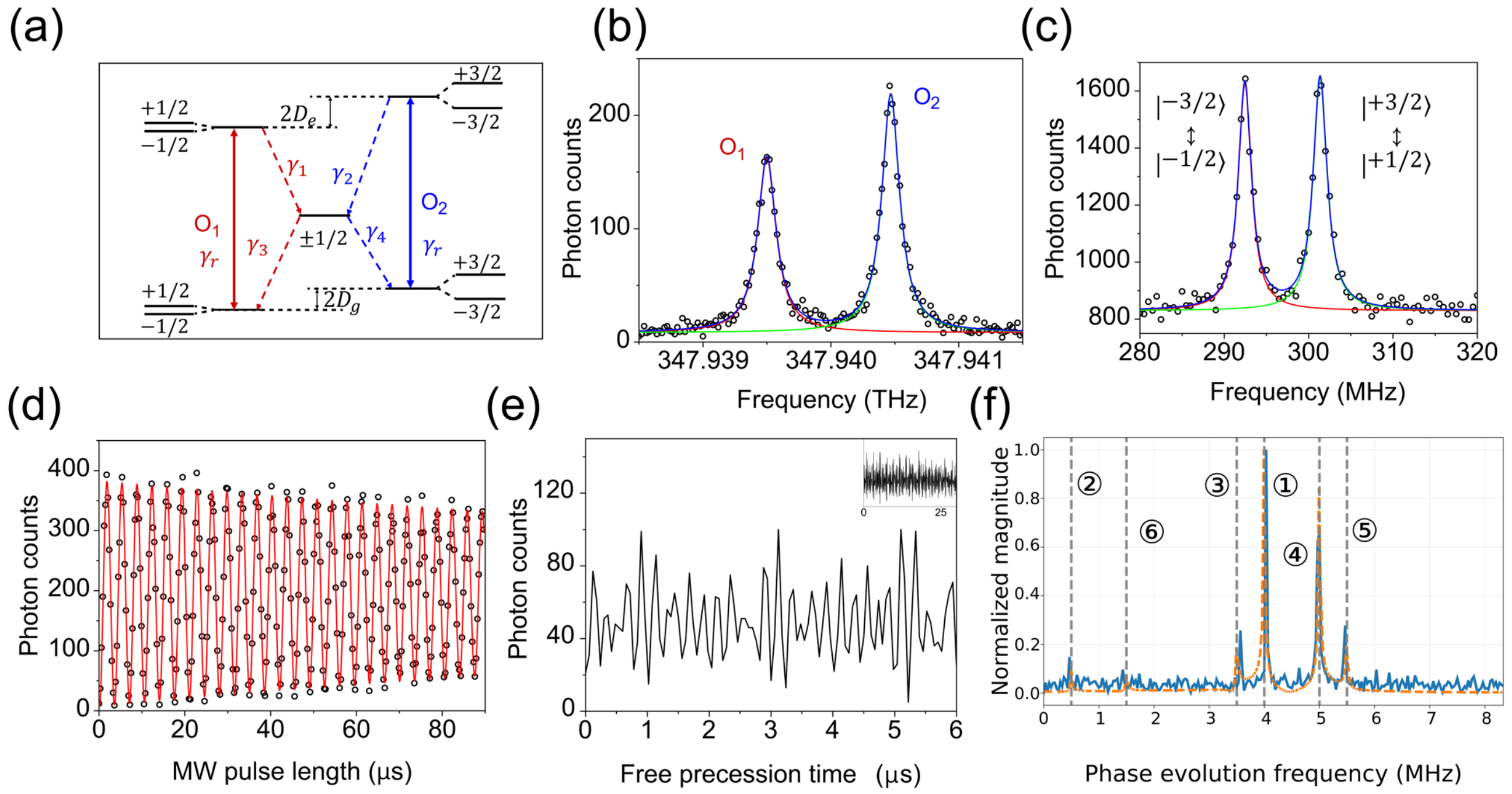

FIG. 1. (a) Level diagram and rate model of the V1 center in 4H-SiC ($O_1$: $m_S = \pm 1/2$; $O_2$: $m_S = \pm 3/2$). $\gamma_r$ is the radiative transition rate, while $\gamma_1$, $\gamma_2$, $\gamma_3$, and $\gamma_4$ are nonradiative spin-dependent intersystem-crossing (ISC) rates [38]. (b) PLE spectrum of the V1 center measured at $T \approx 5$ K and without an external magnetic field. The two peaks correspond to the optical transitions $O_1$ (left) and $O_2$ (right). The solid lines are Lorentzian fits. The excitation laser power in front of the objective lens was 90 nW. (c) cw-ODMR with the laser parked on $O_2$ under a static field of 100 G parallel to the c axis. The two resonances correspond to the $m_S = -3/2 \leftrightarrow m_S = -1/2$ transition (left peak) and the $m_S = +3/2 \leftrightarrow m_S = +1/2$ transition (right peak). The solid lines are Lorentzian fits. The excitation and readout laser power was 7.7 nW. (d) Optically detected Rabi oscillations measured under resonant microwave driving at $m_S = +3/2 \leftrightarrow m_S = +1/2$, with a damped-sine fit (solid line). The plotted data are raw photon counts. The resonant laser power used for both initialization and readout was in the 1–3 nW range throughout the pulse sequence. (e) A close-up of the Ramsey fringes from two $\pi/2$ pulses separated by $\tau$ with detuning $\delta/2\pi = -4$ MHz. The excitation and readout laser power was 40 nW. The inset is the same data in the full range of the free-precession time involved. (f) FFT of the Ramsey fringes (blue solid line), showing multiple frequency components induced by crosstalk between spin transitions. The simulation (orange dashed line) and the numbered frequency positions labeled by ①–⑥ are discussed in Secs. III B–III D.

Figure 1(a) also schematizes the rate model used throughout: a minimal three-manifold scheme with the ground quartet, the excited quartet, and a metastable shelving manifold. Optical driving on $O_1/O_2$ is treated as predominantly $\Delta m_S = 0$ (spin-conserving) transition, while spin-dependent ISC pathways via the metastable state provide the effective spin-flip channel that redistributes population between $|m_S = \pm 1/2\rangle$ and $|m_S = \pm 3/2\rangle$ [38,42]. In practice, $O_1$ ($O_2$) pumping tends to polarize the ground state toward $|m_S = \pm 3/2\rangle$ ($|m_S = \pm 1/2\rangle$) via ISC; this mechanism underlies the initialization protocols used for ODMR [38,39].

To characterize the ground-state spin resonances, we perform continuous-wave ODMR with a constant external magnetic field of 104.80(4) G aligned to the c-axis. With a narrow-linewidth laser fixed on $O_2$, we scan the MW frequency while recording the PSB fluorescence. Under continuous $O_2$ pumping, ISC depletes the $m_S = \pm 3/2$ sublevels and accumulates population in the $m_S = \pm 1/2$ manifold (a mixed state of the two sublevels) [8,12,38]. At MW resonance, the drive induces the $\Delta m_S = \pm 1$ transitions that transfer population from $m_S = \pm 1/2$ into $m_S = \pm 3/2$. Because those sublevels are directly addressed by $O_2$, the photon emission rate rises sharply, yielding high-visibility ODMR contrast. Although near-pure state initialization into a single $m_S$ sublevel is achievable with resonant pumping with broadband MW excitation [12,38], our focus here is crosstalk in the $S = 3/2$ manifold. For this purpose, initialization into the mixed $m_S = \pm 1/2$ manifold is sufficient and operationally simpler, and we adopt it throughout. The resulting spectrum (Fig. 1(c)) shows two distinct lines corresponding to the outer $\Delta m_S = \pm 1$ transitions $m_S = \pm 3/2 \leftrightarrow \pm 1/2$ with the absence of the middle transition $m_S = -1/2 \leftrightarrow +1/2$ in agreement with the well-known $V1$ ground-manifold structure [12].

With the optical and MW resonances identified, we next measure spin Rabi oscillations to calibrate the pulse parameters for coherent control. For high-visibility Rabi measurements, we initialized the spin into a near-pure state of $m_S = +1/2$ using the selective initialization protocol [38]. With a narrow-linewidth laser fixed on $O_2$ and the MW tuned to the resonance at $m_S = +3/2 \leftrightarrow +1/2$, we observe high-contrast optically detected Rabi oscillations [Fig. 1(d)]. A damped-sine fit to the raw data shown in Fig. 1(d) yields a visibility of 96(1)%, which increases to 98(1)% after subtracting a local background measured independently at a nearby emitter-free position at the same focal depth using the same pulse sequence and detection settings. These measurements verify high-fidelity coherent control before the broadband Ramsey experiments presented below, where we quantify crosstalk within the $S = 3/2$ manifold.

We next implement a Ramsey pulse sequence to probe the phase evolution and frequency response of the $S = 3/2$ spin system with $\omega_1/2\pi = 3.125$ MHz . Two 80 ns long $\pi/2$ MW pulses are applied, separated by a variable free-precession interval $\tau$. These nominally square pulses are generated by microwave switching and are therefore not perfectly rectangular. The rise and fall times are about 5 ns and the switching time is about 20 ns, which can modify the effective pulse area and bandwidth relative to the idealized control used in the simulation. The MW was detuned by $-4$ MHz from the chosen outer transition, $m_S = +3/2 \leftrightarrow +1/2$, so that the phase accumulates at 4 MHz during free precession. The second $\pi/2$ pulse maps this accumulated phase to population for optical readout [37]. A representative time-domain trace (Fig. 1(e)) shows fringes that do not reduce to a single-frequency sinusoid; instead, visible multifrequency modulation is observed. To make the frequency content explicit, we compute the FFT of the Ramsey signal (Fig. 1(f)). In addition to the 4 MHz component due to the programmed detuning, the spectrum exhibits additional strong peaks at around 3.5 MHz, 5 MHz and 5.5 MHz and two more small peaks with marginal SNR ratio at around 1 MHz and 1.5 MHz that cannot be accounted for by a two-level model. These features indicate coherent crosstalk among multiple spin transitions within the $S = 3/2$ manifold. Because the trace shown in Fig. 1(e) was chosen to highlight multicomponent beating, it is not suitable for extracting a unique $T_2^*$. Instead, using additional positive-detuning Ramsey traces whose near-single-component character is also consistent with the corresponding Fourier spectra in Figs. 4(b) and 4(d), we conservatively estimate $T_2^* > 30$ μs. The corresponding data and fits are provided in Appendix C. In what follows, we develop a multitransition theoretical framework to quantitatively reproduce the Ramsey spectra and isolate the crosstalk pathways.

### B. Numerical prediction of multilevel Ramsey branches

Motivated by the multifrequency Ramsey response in Figs. 1(e)–1(f), we numerically simulate the Ramsey sequence to preview the frequency branches expected from a four-level $S = 3/2$ manifold under our experimental conditions. All simulations were implemented in Wolfram Mathematica using the SpinDynamica package, which provides a density-matrix/Schrödinger-equation simulation framework for driven spin dynamics[43]. The simulations integrate the time-dependent Schrödinger equation in the rotating frame using the Hamiltonian of Eq. (2) (see Sec. II and Appendix A). We initialize the simulation in an ideal equal mixture confined to the $|\pm 1/2\rangle$ ground manifold, apply a $\pi/2$ pulse with the detuning, allow free precession for a variable interval, and apply a final $\pi/2$ pulse. This captures the dominant experimental preparation while not explicitly including residual population outside the addressed manifold. The measured quantity is the $O_2$ readout channel, modeled by the measurement operator $\widehat{O}_2$, $S(\tau) = \left\langle \psi_f \middle| \widehat{O}_2 \middle| \psi_f \right\rangle$. This effective readout model is intentionally simplified and does not explicitly include finite-duration resonant detection, imperfect spin selectivity, or state-dependent optical pumping and shelving through the intersystem-crossing channels. To isolate spectral content cleanly, we use idealized conditions (long acquisition window and dense sampling) beyond what is available experimentally.

Figure 2 shows the numerical simulations for the detuning-dependent Ramsey spectra obtained by sweeping the MW detuning $\delta/2\pi$ while holding the drive amplitude $\omega_1$ fixed ($\omega_1/2\pi = 5$ MHz) which is intentionally chosen to be a little larger than the experimental value in order to not miss any important features. Under these idealized conditions, the Fourier transform reveals six branches, rather than a single line at $|\delta/2\pi|$ which tells us that the multifrequency features found in the experimental data as in Fig. 1(f) are not artifacts but capture the nature of the ququart. As detailed in Appendix B, these components correspond to pairwise coherences between the four $m_S$ sublevels. The relative amplitudes of these branches are governed by the prepared initial state and by $\delta$ and $\omega_1$ which will be explained in III C. The resulting map reproduces the multipeak structure seen in Fig. 1(f) and motivates a compact physical assignment.

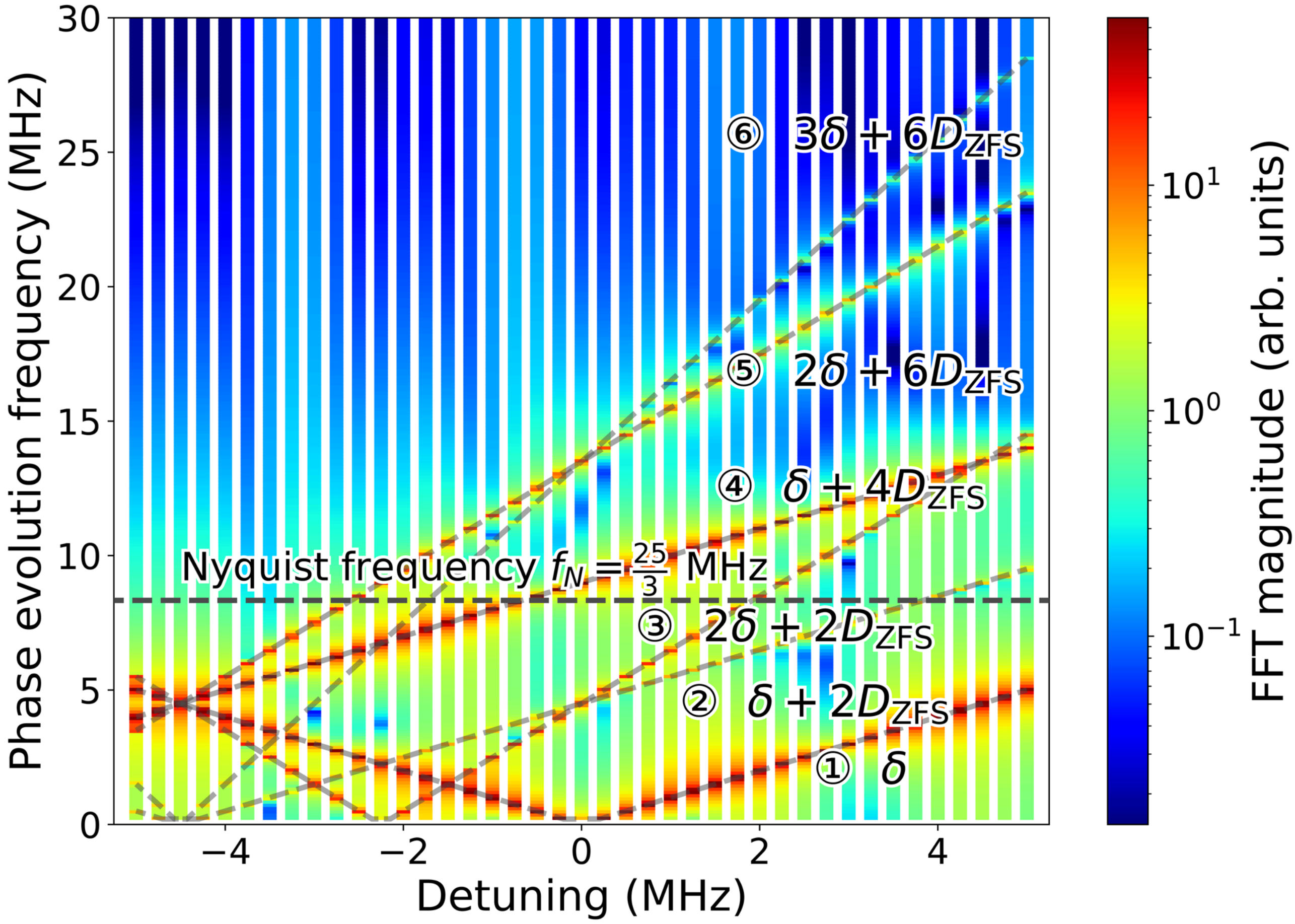

FIG. 2. Simulated Ramsey branch map. FFT magnitude (color) in log scale versus detuning $\delta/2\pi$ for the addressed $m_S = +3/2 \leftrightarrow +1/2$ transition at fixed $\omega_1$. Six branches (①–⑥) are visible; gray dashed red lines are the analytic branch expressions from III C overlaid on the simulation. The spectral branches are ordered and labeled starting from the lowest frequency components generated at small detuning $|\delta|$. The horizontal dashed line at around 8.3 MHz indicates the Nyquist frequency $f_N$ set by the experimental Ramsey sampling interval, above which spectral components are subject to aliasing.

To develop intuition for the multicomponent Ramsey spectra, we visualize the four-level dynamics on two-level Bloch-sphere subspaces extracted from the $S = 3/2$ manifold. While the full evolution unfolds in a four-dimensional Hilbert space, projecting onto selected pairs provides a geometric picture of how detuned $\pi/2$ pulses seed coherences beyond the targeted transition and how free precession redistributes phase. Figure 3 shows simultaneous projections of the same Bloch-vector evolution under a resonant hard pulse followed by free evolution onto three representative neighbor-pair subspaces.

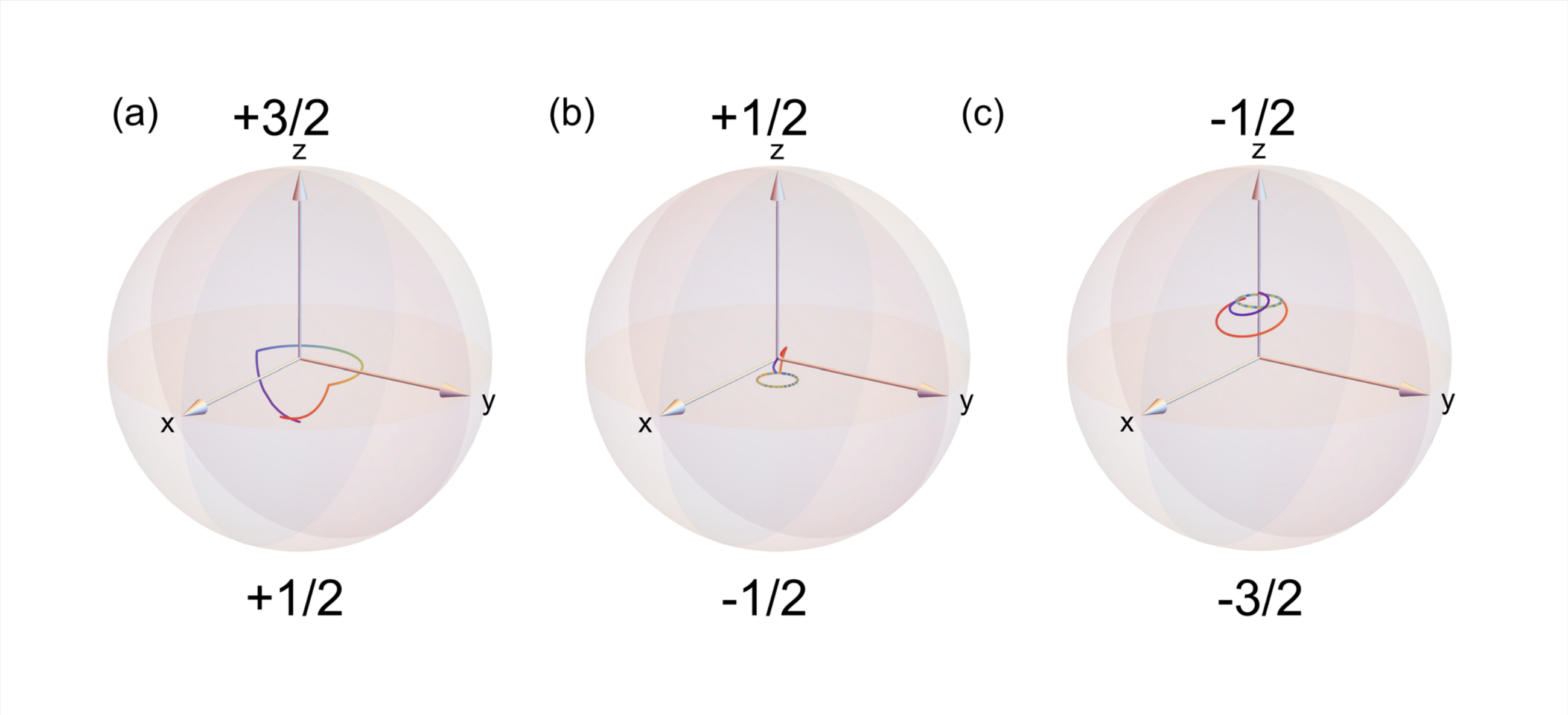

FIG. 3. Bloch-sphere trajectories for a Ramsey sequence driven with resonant hard $\pi/2$ pulses on the outer transition ($m_S = +3/2 \leftrightarrow +1/2$) at $\delta/2\pi = 1$ MHz. The three panels represent simultaneous projections of the same spin-3/2 state evolution during a single Ramsey sequence onto three different two-level subspaces. Panels: (a) subspace $\{+3/2, +1/2\}$, (b) $\{+1/2, -1/2\}$, (c) $\{-1/2, -3/2\}$. The trajectory starts at the purple point and evolves toward the red point, and the surface corresponds to the unit length.

Figure 3(a) depicts the addressed outer pair, $| +3/2\rangle \leftrightarrow | +1/2\rangle$. Starting near the south pole set by the optical initialization, the first $\pi/2$ pulse rotates the Bloch vector to the equator, creating a superposition of the two sublevels. During the subsequent free precession, the Bloch vector winds around the z-axis at a rate set by the detuning $\delta/2\pi$, reproducing the familiar single-pair Ramsey mechanism. If the system were strictly two-level, the trajectory would remain confined to this sphere and the time-domain signal would be close to a single-frequency sinusoid. Panels (b) and (c) show the middle and opposite outer single-quantum pairs, $| +1/2\rangle \leftrightarrow | -1/2\rangle$ and $| -1/2\rangle \leftrightarrow | -3/2\rangle$, respectively. Although these pairs are not directly resonant with the applied drive, the finite spectral bandwidth of the short pulses (e.g., 50 ns in this simulation) and the multilevel couplings induce weak, off-resonant rotations that generate residual coherences in these subspaces. As a result, their Bloch vectors depart from the poles and trace small but coherent trajectories. These secondary motions seed additional Fourier components in the Ramsey signal once the accumulated phases are projected back to populations by the second $\pi/2$ pulse and read out on $O_2$. Thus, even with a drive nominally targeting a single outer transition, coherence is not confined to one two-level manifold. In what follows we refer to these off-resonant pathways into nonaddressed pairs as spectator transitions. Taken together, the three Bloch-sphere projections in Fig. 3 provide a dynamical picture consistent with the numerical branch map: the Ramsey sequence excites coherences in multiple neighbor pairs, leading to a spectrum with several lines rather than a single detuning component. This visualization is intended as an intuitive guide; a quantitative assignment of branch frequencies and their weights in terms of pairwise coherences is given in Appendix B and used in the analytical treatment that follows.

## C. Analytical frequency set and mapping

Guided by the numerical preview in Sec. III B, we now state the analytic frequency set for the Ramsey response of the $S = 3/2$ manifold. In the rotating frame (Eq. (2)), the free-evolution phases are governed by the level energies $E_m$ and the Ramsey signal takes the form (see Appendix B for details)

$$S(\tau) = S_0 + \sum_{m<n} 2|X_{mn}| \cos(\Omega_{mn}\tau - \arg X_{mn}) , \quad (6)$$

where each of the six coherences $(m, n)$ in the $S = 3/2$ manifold contributes a cosine at a frequency $\Omega_{mn}/2\pi$ and with a weight $| X_{mn} |$ set by the prepared state and the $\pi/2$ pulse matrix (see Appendix B). The underlying frequencies $\{\Omega_{mn}/2\pi\}$ are the six pairwise energy splittings of the rotating-frame Hamiltonian as in Eq. (5) and are listed and labeled in Fig. 2 as branches ①–⑥ (ordered by increasing frequency at small $|\delta/2\pi|$). That labeling is used throughout this section and in Fig. 4. For clarity, except for the branch associated with the addressed outer single-quantum pair with $\Omega_{+1/2,+3/2} = \delta$, all the remaining components originating from off-resonant pairs are 'spectator-transition branches' that acquire coherence during the sequence. Because the data (and the simulations in Fig. 4) are sampled with a sampling rate $f_s = 1/(60\text{ ns})$, only the first Nyquist zone $[0, f_N]$ with the Nyquist frequency $f_N \equiv f_s/2$ is observable as in Figs. 4(a) and 4(b) [44,45]. As an example, the unaliased phase evolution frequency of the branch ③ is $\Omega_{-1/2,+3/2}/2\pi = (2\delta + 2D_{gs})/2\pi$ when $\delta/2\pi \le (f_N - 2D_{gs}/2\pi)/2$ but the aliased frequency becomes $\Omega_{-1/2,+3/2}/2\pi = 2f_N - (2\delta + 2D_{gs})/2\pi$ when $\delta/2\pi > (f_N - 2D_{gs}/2\pi)/2$ due to the Nyquist folding (mirror about $f_N$) as in Figs. 4(a) and 4(b). Another example is the frequency of the branch ⑥ $\Omega_{-3/2,+3/2}/2\pi = (3\delta + 6D_{gs})/2\pi$, whose unaliased frequency does not fall within the first Nyquist zone even at the zero detuning. In the end, it shows the folding three times in the considered frequency range in Figs. 4(a) and 4(b); $\Omega_{-3/2,+3/2}/2\pi = 2f_N - (3\delta + 6D_{gs})/2\pi$ when $\delta/2\pi \le (2f_N - 6D_{gs}/2\pi)/3$, $-2f_N + (3\delta + 6D_{gs})/2\pi$ when $(2f_N - 6D_{gs}/2\pi)/3 < \delta/2\pi \le f_N - 2D_{gs}/2\pi$, and $4f_N - (3\delta + 6D_{gs})/2\pi$ when $\delta/2\pi > f_N - 2D_{gs}/2\pi$ [45]. The set of six observed frequencies in the simulations of Figs. 4(a) and 4(b) is precisely the image of the unaliased $\{\Omega_{mn}/2\pi\}$ under this map; the branch identities (①–⑥) are preserved continuously through the folds. The positions are fully determined by the level spacings (Sec. II), while the relative amplitudes are set by the prepared initial state and by $\delta$ and $\omega_1$ via the pulse matrix (Appendix B). This frequency mapping explains the multiridge structure of the detuning sweep and the apparent "reflections" where lines cross the Nyquist boundary.

## D. Quantitative comparison with experiment

We now benchmark the model against the Ramsey measurements at the various detuning frequencies used experimentally. Figures 4(c)–4(d) overlay the experimental FFT with the numerical spectrum computed under the same acquisition window and sampling as the data, and with the analytic frequency ticks (vertical markers) for branches ①–⑥. The only inputs are the calibrated parameters from Sec. III A (resonant frequencies, $\pi/2$ pulse), the fixed drive amplitude $\omega_1/2\pi$=3.125 MHz, and the working detuning varies between $\delta/2\pi = -4$ and 3 MHz; no additional parameter is introduced other than an overall vertical scale.

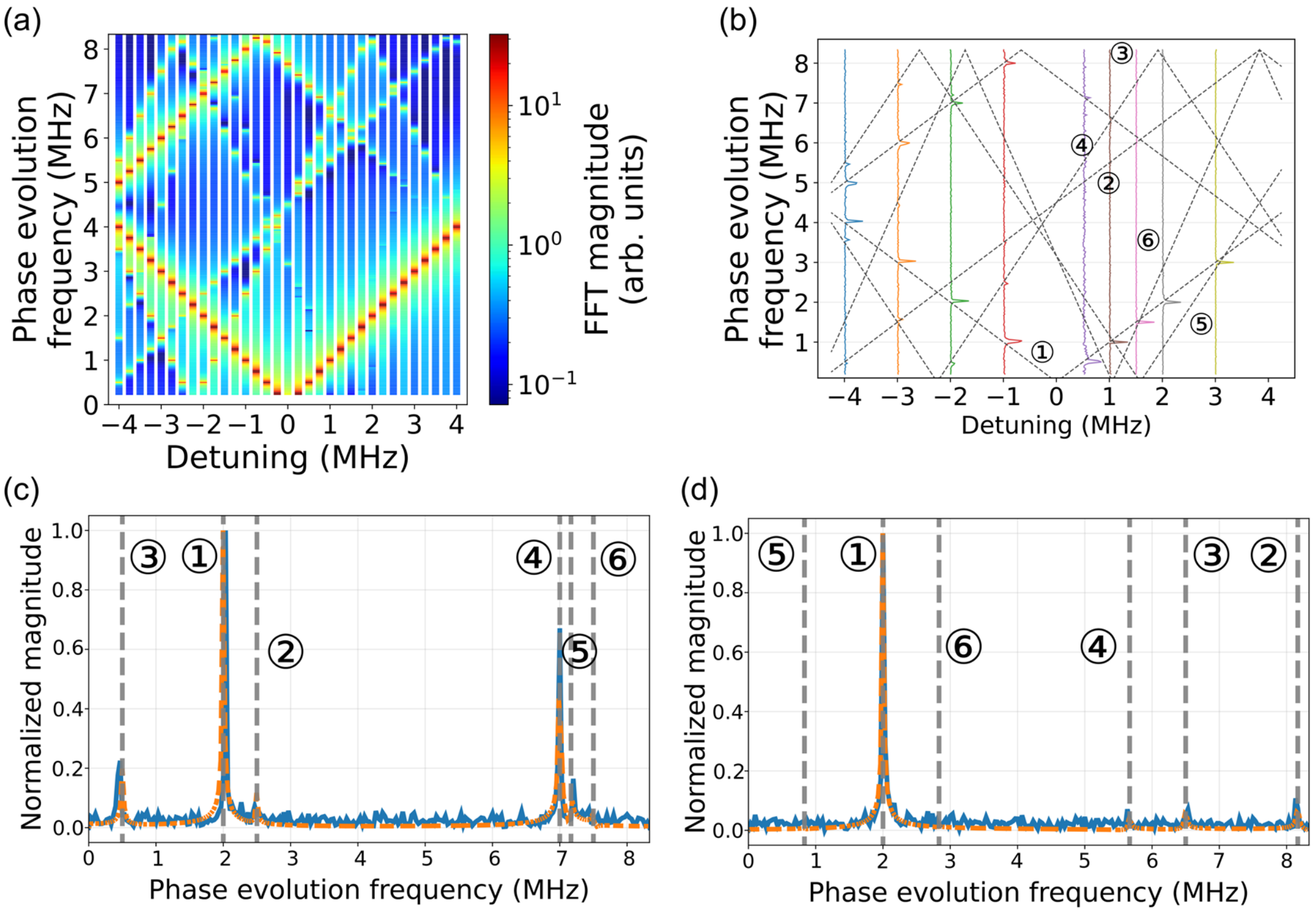

FIG. 4. (a) Detuning sweep under experimental sampling. Fourier amplitude from numerical simulation in log scale as a function of detuning $\delta/2\pi$ using the same experimental parameters including the fixed drive $\omega_1$, time step $\Delta\tau$ and window as the experiment. (b) Experimental Ramsey data, shown as the normalized magnitude of the extracted phase-evolution frequencies, vertically offset and plotted as a function of detuning. (c) Experimental (blue solid lines) and simulated (orange dashed lines) spectra at a detuning of -2 MHz, (d) +2 MHz. For (c), and (d), the six branches ①–⑥ found by analytic solutions are indicated as well.

Across the used detunings, the peak positions in the FFT follow the analytic frequency set after Nyquist folding (cf. Sec. III C): components that cross the Nyquist boundary appear mirrored about $f_N$, and the data peaks track these folded ticks within the frequency resolution set by the acquisition window. To compare the branch frequencies in Figs. 4(c) and 4(d), we extracted the center of each resolved Fourier peak in both the experimental and simulated Ramsey spectra by fitting the central part of the peak with a Lorentzian function. We use this here as a practical peak-position estimator. For a weakly damped, nearly single-frequency Ramsey component, the Fourier peak becomes narrow and approaches a delta-like line in the long-time limit. In the present case, however, the spectra are finite in duration and contain multiple nearby components, so the Lorentzian fit should be understood as an empirical extraction method rather than an exact line-shape model. With this procedure, the dominant addressed branch is reproduced within a few to a few tens of kHz over the analyzed detuning range. This one-to-one agreement confirms the six-branch structure predicted for a spin-3/2 Ramsey sequence and validates the branch labels defined in Fig. 2 and used throughout Fig. 4. The remaining frequency mismatch is attributed mainly to slight peak pulling caused by overlap between nearby Fourier components in the finite-time multicomponent spectra, together with small residual imperfections of the control Hamiltonian that are not explicitly included in the minimal model, such as slight magnetic-field misalignment with respect to the defect-spin quantization axis.

The relative amplitudes of the branches are reproduced qualitatively by the numerics: components associated with the addressed outer single-quantum pair dominate, while additional lines, originating from coherences in other pair subspaces, appear with smaller but repeatable weights. Remaining differences in absolute height are expected because the simulated initial state is an ideal equal mixture confined to the $|\pm 1/2\rangle$ manifold, whereas the experimental $O_2$-based optical preparation is not expected to realize this manifold perfectly. In addition, the exact pulse area and bandwidth are affected by the nonideal switched microwave pulse shape, and the simulated readout uses a simplified effective conversion from the final spin populations to the measured optical signal without explicitly modeling finite-duration resonant detection, imperfect spin selectivity, and state-dependent optical pumping and shelving through the intersystem-crossing channels. Small uncertainty in the effective drive amplitude used in the simulation can further contribute to these amplitude differences. Importantly, the set and locations of the peaks require no free parameters beyond those already calibrated.

Taken together, Figure 4 shows that the experiment, the numerical time-domain propagation, and the analytical mapping all point to the same picture: the Ramsey response of the $V1$ center's $S = 3/2$ manifold comprises six well-defined spectral branches, with positions fixed by the level spacings and detuning, and weights governed by the prepared state and pulse parameters. To further clarify the drive dependence of the spectator branches, we performed an additional numerical analysis at a representative detuning of $\delta/2\pi = -1$ MHz while varying the π/2-pulse duration for the addressed transition. Short hard pulses produce strong spectator-induced side branches, whereas longer soft pulses progressively suppress them by improving spectral selectivity. The corresponding spectra are provided in Appendix D.

## IV. CONCLUSIONS

We have shown that Ramsey interferometry on the V1 center's $S = 3/2$ manifold yields a deterministic six-component spectrum that reflects pairwise coherences across the four spin sublevels. Starting from spin-selective resonant optics ($O_1/O_2$) and coherent control, we measured high-visibility Ramsey traces whose Fourier spectra contain multiple lines. A compact analytical description connects each observed frequency $\Omega_{mn}$ to a specific sublevel pair $(m, n)$ and assigns its weight to the complex amplitude $|X_{mn}|$ determined by the prepared state and the two $\pi/2$ pulses. Numerical propagation under experimental sampling reproduces the full detuning map, including Nyquist-folded features, and the experimental FFT peaks coincide with the analytic frequency ticks without additional fitting for peak positions.

Short finite-bandwidth $\pi/2$ pulses seed coherences not only in the addressed transition but also via spectator transitions, off-resonant sublevel pairs that acquire small yet coherent amplitudes during the sequence. The second $\pi/2$ pulse projects all accumulated phases onto the $O_2$ readout, so the Ramsey signal becomes a weighted superposition of six pair coherences. In this regime, crosstalk is intrinsic, not a minor imperfection of the measurement.

Practically, the framework provides clear guidance for control: branch positions are fixed by the level spacings and detuning, while branch weights $| X_{mn} |$ can be tuned through initialization and the pulse parameters $(\delta, \omega_1, \tau_p)$. In the present V1 system, the observed crosstalk arises because short hard $\pi/2$ pulses have an effective bandwidth comparable to the spacing between neighboring transitions, so spectator subspaces are coherently excited in addition to the nominally addressed outer transition. This produces coherent leakage out of the intended control manifold, redistributes spectral weight into additional Ramsey branches, and reduces transition selectivity. In this regime, a purely two-level interpretation can misestimate quantities such as detuning, contrast, and dephasing, and can bias calibration or state/process-estimation routines if the spectator coherences are ignored. Crosstalk can be mitigated most directly by replacing hard pulses with spectrally more selective soft pulses: lowering the microwave power increases the $\pi/2$-pulse duration, narrows the effective control bandwidth, and suppresses off-resonant excitation of spectator transitions, whereas shorter hard pulses enhance spectator excitation and the multicomponent Ramsey response. More generally, the same control map in terms of $| X_{mn} |$ and $(\delta, \omega_1, \tau_p)$ can be used to choose pulse and detuning settings that suppress unwanted spectator branches, or conversely to exploit them for quantum state/process tomography or other forms of state estimation based on the measured amplitudes and phases.

Beyond the specific defect studied here, the analysis and mapping generalize to multilevel qudits: in any spin manifold, Ramsey frequency content is determined by energy differences between sublevels, so the approach immediately predicts which spectral components can appear and how they fold under finite sampling. In practice, however, the clear observation of spectator-state dynamics relies on microwave pulses whose effective bandwidth is comparable to the spacing between nearby transitions. Thus, while the underlying mechanism is general to multilevel qudits under finite-bandwidth control, the present hard-pulse protocol is most readily applicable to qudit systems whose relevant level spacings lie within the bandwidth achievable with realistic microwave field strengths. Looking ahead, the methodology invites systematic exploration of pulse parameter regimes that minimize or shape crosstalk and motivates studies of how spectator transitions impact Hahn-echo and broader dynamical-decoupling sequences. It also suggests that the $| X_{mn} |$ structure can be used to design tomography and estimation protocols that reconstruct either the prepared state or a pulse-induced unitary in situ from the multicomponent Ramsey response. Together, these elements provide a quantitative and portable description of multilevel Ramsey physics, enabling both more robust control and new metrological opportunities in qudit platforms.

## ACKNOWLEDGMENTS

J. U. H. acknowledges support from the Swedish Research Council (VR) (Grant No. 2020-05444) and the European Union under Horizon Europe for the project SPINUS (Grant No. 101135699). N. T. S. acknowledges support from the European Union under Horizon Europe for projects QuSPARC (Grant No. 101186889), QRC-4-ESP (Grant No. 101129663), and QUEST (Grant No. 101156088), and from Vinnova (Grants No. 2024-00461, No. 2025-03848, and No. 2024-03597), the latter supporting the QUASIC project within QSIP. J. U. H. and N. T. S. acknowledge support from the Knut and Alice Wallenberg Foundation (Grant No. KAW 2018.0071). J. K. S. acknowledges that this work was supported by the KOMAC operation fund of KAERI by MSIT (Ministry of Science and ICT). This work was supported by the Institute of Information & Communications Technology Planning & Evaluation (IITP) (Grants No. II220198, No. RS-2025-02306096), and the National Research Foundation of Korea (NRF) (Grants No. 2021R1A2C2006904, No. 2022M3H3A106307411) funded by the Korean government (Ministry of Science and ICT).

## DATA AVAILABILITY

There are no publicly available research data or software supporting this manuscript. Requests for further information or data should be sent to the authors.

## APPENDIX A: DETAILS FOR SEC. II B: SPIN HAMILTONIAN

### 1. Basis, symbols, and detuning convention

We use the ordered $\hat{S}_z$ basis $\{|+3/2\rangle, |+1/2\rangle, |-1/2\rangle, |-3/2\rangle\}$, set $\hbar = 1$, and parametrize the axial ZFS of a spin-3/2 manifold with ZFS $= 2D_{gs}/2\pi$. A constant offset $-5D_{gs}/4$ shifts all eigenenergies equally and does not affect transition frequencies or observable dynamics apart from an overall global phase. We follow the Sec. II B detuning convention $\delta \equiv \omega - (\omega_0 + 2D_{gs})$ so that $\delta > 0$ corresponds to a drive set above the $|+3/2\rangle \leftrightarrow |+1/2\rangle$ resonance at $\omega_0 + 2D_{gs}$.

### 2. Explicit 4 × 4 matrix of the rotating-frame Hamiltonian

In the ordered basis above, the spin-3/2 operators are

$$S_z = \begin{pmatrix} \frac{3}{2} & 0 & 0 & 0 \\ 0 & \frac{1}{2} & 0 & 0 \\ 0 & 0 & -\frac{1}{2} & 0 \\ 0 & 0 & 0 & -\frac{3}{2} \end{pmatrix}, \quad S_x = \frac{1}{2}\begin{pmatrix} 0 & \sqrt{3} & 0 & 0 \\ \sqrt{3} & 0 & 2 & 0 \\ 0 & 2 & 0 & \sqrt{3} \\ 0 & 0 & \sqrt{3} & 0 \end{pmatrix}, \quad S_y = \frac{1}{2i}\begin{pmatrix} 0 & \sqrt{3} & 0 & 0 \\ -\sqrt{3} & 0 & 2 & 0 \\ 0 & -2 & 0 & \sqrt{3} \\ 0 & 0 & -\sqrt{3} & 0 \end{pmatrix}. \quad \text{(A1)}$$

These matrices explain the unequal transverse couplings: $|\pm3/2\rangle \leftrightarrow |\pm1/2\rangle$ carry $\sqrt{3}/2$ , whereas $|+1/2\rangle \leftrightarrow |-1/2\rangle$ carries 1. Hence, from Eq. (2) in the main text

$$\hat{H}_{rot} = \begin{pmatrix} D_{gs} + \frac{3}{2}(\omega_0 - \omega) & \frac{\sqrt{3}}{2}\omega_1 & 0 & 0 \\ \frac{\sqrt{3}}{2}\omega_1 & -D_{gs} + \frac{1}{2}(\omega_0 - \omega) & \omega_1 & 0 \\ 0 & \omega_1 & -D_{gs} - \frac{1}{2}(\omega_0 - \omega) & \frac{\sqrt{3}}{2}\omega_1 \\ 0 & 0 & \frac{\sqrt{3}}{2}\omega_1 & D_{gs} - \frac{3}{2}(\omega_0 - \omega) \end{pmatrix}. \tag{A2}$$

### 3. Free-evolution spectrum and the six Ramsey frequencies

With the drive off ($\omega_1 = 0$) during the free-precession, the free Hamiltonian is diagonal:

$$\hat{H}_{\text{free}} = -(\delta + 2D_{gs})\hat{S}_z + D_{gs}\left(\hat{S}_z^2 - \frac{5}{4}\hat{I}\right), \tag{A3}$$

with the eigenvalues

$$E_{-3/2} = \frac{3}{2}\delta + 4D_{gs}, E_{-1/2} = \frac{\delta}{2}, E_{+1/2} = -\frac{\delta}{2} - 2D_{gs}, E_{+3/2} = -\frac{3}{2}\delta - 2D_{gs}. \tag{A4}$$

All pairwise differences $\Omega_{mn} = |E_m - E_n|$ yield the six detuning-dispersive components observed in Figs. 2 and 4,

$$\{\Omega\} = \{\delta, \delta + 2D_{gs}, \delta + 4D_{gs}, 2\delta + 2D_{gs}, 2\delta + 6D_{gs}, 3\delta + 6D_{gs}\}. \tag{A5}$$

These frequency components will also be directly derived from the observables, $S(\tau) = \left\langle \psi_f \middle| \hat{O}_2 \middle| \psi_f \right\rangle$, as in Appendix B.

## APPENDIX B: DETAILS FOR SEC. II C: RAMSEY SEQUENCE AND FREQUENCY DECOMPOSITION

### 1. Sequence, initial state, and measurement

We model the Ramsey sequence by

$$\hat{U}_{Ramsey}(\tau) = \hat{R}_x\left(\frac{\pi}{2}\right) e^{-i\hat{H}_{free}\tau} \hat{R}_x\left(\frac{\pi}{2}\right), \tag{B1}$$

where $H_{free}$ is the free-evolution Hamiltonian in Appendix A, $\hat{R}_x(\pi/2)$ are the first/second $\pi/2$ pulses (general 4×4 unitaries) of which relative MW phases are zero. $\tau$ is free-precession time in which $\omega_1$ is off. The initial state $\rho_0$ is set to be a pure state $|+1/2\rangle$. Although the V1 center in this work is initialized into the equal statistical mixture of $|\pm1/2\rangle$ [38], this assumption will allow us to simplify the problem so that we can find the solution for the frequency components of the time-evolution by the used Ramsey sequence. Note that our interest is to understand the origin of the phase evolution frequency and its relation to the detuning.

We read out the signal

$$S(\tau) = \mathrm{Tr}\big[\hat{O}_2 \hat{U}_R(\tau) \rho_0 \hat{U}_R^{\dagger}(\tau)\big], \tag{B2}$$

with the default choice $\hat{O}_2 = | +3/2\rangle\langle +3/2 | + | -3/2\rangle\langle -3/2 |$, consistent with the experimental observable corresponding to the $O_2$ transition. The other observable $\hat{O}_1 = | +1/2\rangle\langle +1/2 | + | -1/2\rangle\langle -1/2 |$ only reweights amplitudes but does not create new frequency branches since $\langle O_1\rangle + \langle O_2\rangle = 1$.

## 2. Free-evolution spectrum and the six branches

As explained in Sec. A 3, the free-evolution eigenenergies in the rotating frame read four different values so that observable frequencies are all pairwise differences $\Omega_{mn}$ (six in total). Using the detuning and ZFS conventions (ZFS=$2D_{gs}$) one obtains the following six linear branches (slope–intercept form with respect to $\delta$) as mentioned in Sec. A 3 and summarized in Table I. These six branches are exactly the detuning-dispersive components used to annotate Figs. 2 and 4 in the main text.

TABLE I. During the Ramsey free evolution, the pairwise differences $\Omega_{mn} = |E_m - E_n|$ of the diagonal energies generate six linear branches as functions of detuning $\delta$. This table lists the branch labels (①–⑥) used in Figs. 2 and 4, the analytic forms of $\Omega_{mn}(\delta)$, their slope-intercept pairs, and the $(m, n)$ pairs in the ordered basis $\{|+3/2\rangle, |+1/2\rangle, |-1/2\rangle, |-3/2\rangle\}$.

| Branch | Pair $(m, n)$ | (slope, intercept) | $\Omega_{\mathrm{mn}}(\delta)$ |
|---|---|---|---|
| ① | $(+\frac{1}{2}, +\frac{3}{2})$ | $(1, 0)$ | $\delta$ |
| ② | $:(-\frac{1}{2}, +\frac{1}{2})$ | $(1, 2D_{gs})$ | $\delta + 2D_{gs}$ |
| ③ | $:(-\frac{1}{2}, +\frac{3}{2})$ | $(2, 2D_{gs})$ | $2\delta + 2D_{gs}$ |
| ④ | $:(-\frac{1}{2}, -\frac{3}{2})$ | $(1, 4D_{gs})$ | $\delta + 4D_{gs}$ |
| ⑤ | $:(-\frac{3}{2}, +\frac{1}{2})$ | $(2, 6D_{gs})$ | $2\delta + 6D_{gs}$ |
| ⑥ | $:(-\frac{3}{2}, +\frac{3}{2})$ | $(3, 6D_{gs})$ | $3\delta + 6D_{gs}$ |

## 3. From the first $\pi/2$ pulse to the Ramsey signal

A general (detuned, finite-bandwidth) first $\pi/2$ pulse makes all four components nonzero. We therefore do not write the pulse matrix explicitly here and simply denote the post-pulse state as

$$|\psi_1\rangle \equiv \begin{pmatrix} c_1 e^{i\Delta_1} \\ c_2 e^{i\Delta_2} \\ c_3 e^{i\Delta_3} \\ c_4 e^{i\Delta_4} \end{pmatrix}, \tag{B3}$$

with real $c_k \geq 0$ and pulse-accumulated phases $\Delta_k$. During the free-precession interval, since $H_{free}$ is diagonal as explained in Eq. (A3), the state right before the second $\pi/2$ pulse, i.e. $e^{-i\hat{H}_{free}\tau} \mid \psi_1\rangle$ is, using the eigenvalues in Eq. (A4) and dropping a global phase term,

$$\begin{pmatrix} c_1 e^{i\Delta_1} e^{i\delta\tau} \\ c_2 e^{i\Delta_2} \\ c_3 e^{i\Delta_3} e^{-i(\delta+2D_{gs})\tau} \\ c_4 e^{i\Delta_4} e^{-i(2\delta+6D_{gs})\tau} \end{pmatrix}. \tag{B4}$$

To find the explicit form of the ququart state right after the second $\pi/2$ pulse, we investigate the full 4×4 matrix form. To expose the structure of the second $\pi/2$ pulse, it is useful to show the ideal hard-resonant rotation about x-axis for spin-3/2. During a short and high-power MW pulse, of which the timescale is $\tau_p$, we work in the rotating frame and neglect the internal/free Hamiltonian $H_{free}$ while the pulse is on; the control Hamiltonian dominates. "Hard" means the MW field amplitude $\omega_1$ is large and the pulse is brief so that evolution under $H_{free}$ is negligible over the pulse $\left(\tau_p \ll 1/|H_{\text{free}}|\right)$. "Resonant" means the drive is on resonance with the targeted transitions (detuning $\delta = 0$), so in the rotating-wave approximation the effective Hamiltonian during the pulse is $\omega_1 \hat{S}_x$. Consequently, the pulse implements a pure spin-3/2 rotation about x-axis:

$$\hat{U}_p(\omega_1\tau_p) = e^{-i\omega_1\tau_p\hat{S}_x} \equiv \hat{R}_x(\omega_1\tau_p). \tag{B5}$$

Note that these are standard magnetic resonance control assumptions: on-resonance RWA plus the "hard-pulse" approximation that internal terms (e.g., Zeeman offsets) can be ignored during $\tau_p$. During a pulse of duration $\tau_p$ we neglect the zero-field splitting $D_{gs}$ (hard-pulse approximation) but keep a static detuning $\delta$ to keep the generality. With Rabi amplitude $\omega_1$ the effective Hamiltonian is

$$\hat{H}_p \simeq -\delta\hat{S}_z + \omega_1\hat{S}_x = \Omega_{\text{Rabi}}(\hat{S}_z\cos\beta - \hat{S}_x\sin\beta) = \Omega_{\text{Rabi}} e^{+i\beta\hat{S}_y}\hat{S}_z e^{-i\beta\hat{S}_y}, \tag{B6}$$

where the effective Rabi rate $\Omega_{Rabi}$ and tilt angle $\beta$ are defined as

$$\begin{aligned} &\Omega_{\text{Rabi}} = \sqrt{\delta^2 + \omega_1^2}, \qquad \cos\beta = -\frac{\delta}{\Omega_{\text{Rabi}}}, \\ &\sin\beta = -\frac{\omega_1}{\Omega_{\text{Rabi}}}, \qquad -\pi < \beta < 0 \ (\omega_1 > 0). \end{aligned} \tag{B7}$$

Then, the time evolution operator is

$$\hat{U}_p(\tau_p) = e^{-i(-\delta\hat{S}_z + \omega_1\hat{S}_x)\tau_p} = e^{+i\beta\hat{S}_y} e^{-i\Omega_{\text{Rabi}}\tau_p\hat{S}_z} e^{-i\beta\hat{S}_y}, \tag{B8}$$

where we used the similarity rule for operator exponentials, $e^{\hat{U}^\dagger\hat{A}\hat{U}} = \hat{U}^\dagger e^{\hat{A}}\hat{U}$ (with $\hat{U} = e^{-i\beta\hat{S}_y}$, $\hat{A} = -i\Omega_{\text{Rabi}}\tau_p\hat{S}_z$). Note that the final form in Eq. (B8) is not the standard $z - y - z$ Euler decomposition. Here, since the Wigner D-matrix is defined as $\hat{D}^{(j)}(\alpha,\beta,\gamma) = e^{-i\alpha\hat{S}_z}e^{-i\beta\hat{S}_y}e^{-i\gamma\hat{S}_z}$ with $j = 3/2$, we express $U_p(\tau_p)$ using the products of Wigner D matrices [46] for three rotations as

$$\hat{U}_p(\tau_p) = \hat{D}^{(3/2)}(0,-\beta,0)\hat{D}^{(3/2)}(\Omega_{\text{Rabi}}\tau_p,0,0)\hat{D}^{(3/2)}(0,\beta,0)$$

$$= d^{(3/2)}(-\beta)\text{diag}\left(e^{-\frac{3i}{2}\Omega_{\text{Rabi}}\tau_p}, e^{-\frac{i}{2}\Omega_{\text{Rabi}}\tau_p}, e^{+\frac{i}{2}\Omega_{\text{Rabi}}\tau_p}, e^{+\frac{3i}{2}\Omega_{\text{Rabi}}\tau_p}\right) d^{(3/2)}(\beta), \tag{B9}$$

where $d^{3/2}(\beta)$ is a Wigner small-d matrix for $j = 3/2$ and rotating angle $\beta$, i.e.,

$$d^{\left(\frac{3}{2}\right)}(\beta) \equiv \langle m'|e^{-i\beta\hat{S}_y}|m\rangle = \begin{pmatrix} c^3 & -\sqrt{3}c^2s & \sqrt{3}cs^2 & -s^3 \\ \sqrt{3}c^2s & c(c^2-2s^2) & (s^2-2c^2)s & \sqrt{3}cs^2 \\ \sqrt{3}cs^2 & (2c^2-s^2)s & c(c^2-2s^2) & -\sqrt{3}c^2s \\ s^3 & \sqrt{3}cs^2 & \sqrt{3}c^2s & c^3 \end{pmatrix}. \tag{B10}$$

where $c \equiv \cos(\beta/2)$ and $s \equiv \sin(\beta/2)$, with $2c^2 - 1 = \cos\beta$. For generic nonzero drive and pulse parameters, Eqs. (B9) and (B10) show that every element of $U_p$ is a nontrivial linear combination of four complex phases with real weights; hence, no parameter-independent structural zeros appear. To express the analytical form of the full $4 \times 4$ matrix of $\hat{U}_p(\tau_p)$, we additionally define new parameters, $\theta \equiv \Omega_{\text{Rabi}}\tau_p$, $q \equiv e^{-i\theta/2}$, $p \equiv e^{-3i\theta/2}$, then using Eqs. (B9) and (B10), we find the matrix elements of $\hat{U}_p(\tau_p)$ as below,

$$U_{+3/2,+3/2} = pc^6 + 3qc^4s^2 + 3q^*c^2s^4 + p^*s^6$$

$$U_{+3/2,+1/2} = \sqrt{3}cs[-pc^4 + qc^2(c^2-2s^2) + q^*s^2(2c^2-s^2) + p^*s^4]$$

$$U_{+3/2,-1/2} = \sqrt{3}c^2s^2[pc^2 + q(s^2-2c^2) + q^*(c^2-2s^2) + s^2p^*]$$

$$U_{+3/2,-3/2} = c^3s^3(-p + 3q - 3q^* + p^*)$$

$$U_{+1/2,+1/2} = 3pc^4s^2 + qc^2(c^2-2s^2)^2 + q^*s^2(2c^2-s^2)^2 + 3p^*c^2s^4$$

$$U_{+1/2,-1/2} = cs(c^2-2s^2)(2c^2-s^2)(q^*-q) + 3c^3s^3(p^*-p)$$

$$U_{+1/2,-3/2} = \sqrt{3}c^2s^2[p^*c^2 + q^*(s^2-2c^2) + q(c^2-2s^2) + s^2p]$$

$$U_{-1/2,-1/2} = 3p^*c^4s^2 + q^*c^2(c^2-2s^2)^2 + qs^2(2c^2-s^2)^2 + 3pc^2s^4$$

$$U_{-1/2,-3/2} = \sqrt{3}cs[p^*c^4 - q^*c^2(c^2-2s^2) - qs^2(2c^2-s^2) - ps^4]$$

$$U_{-3/2,-3/2} = p^*c^6 + 3q^*c^4s^2 + 3qc^2s^4 + ps^6 \tag{B11}$$

We find that $\hat{U}_p^T = \hat{U}_p$ and $U_{mn} = (-1)^{m-n}U^*_{-m,-n}$, where $m, n \in \{+3/2, +1/2, -1/2, -3/2\}$ and $m - n$ is an integer. Then, the observable becomes, according to Eq. (B2),

$$S(\tau) = \langle\psi_f|\hat{O}_2|\psi_f\rangle = c_0 + \textstyle\sum_{m<n} 2\,|X_{mn}|\cos[\Omega_{mn}\tau - \arg X_{mn}], \tag{B12}$$

where

$$c_0 = \sum_m |U_{m,+1/2}|^2 \left(|U_{+3/2,m}|^2 + |U_{-3/2,m}|^2\right),$$

$$X_{mn} = U_{m,+1/2}U^*_{n,+1/2}\left(U_{+3/2,m}U^*_{+3/2,n} + U_{-3/2,m}U^*_{-3/2,n}\right), \tag{B13}$$

and $U_{ab} \equiv \langle a | \hat{U}_p | b \rangle$.

At this point, the main objective of Appendix B is achieved by finding the cosine term in Eq. (B12). Because the final $\pi/2$ pulse is a generic unitary acting on the four-level manifold, its matrix elements are in general nonzero, and multiplying $U_p$ by the freely evolving state inevitably mixes the pairwise phases accumulated between all sublevel pairs. Consequently, the measured signal can always be organized as a sum of cosine terms oscillating at the six pairwise splittings $\Omega_{mn}$. Finding the explicit form of each coefficient $X_{mn}$ is not strictly required for establishing the frequency set; the expressions are included to show explicitly that the second $\pi/2$ pulse is nontrivial and to obtain a closed-form expression under the hard-pulse approximation. However, this approximation also introduces an unavoidable inconsistency in the weights $X_{mn}$. A representative example is the $(m, n) = (-1/2, +1/2)$ contribution associated with the $|-1/2\rangle \leftrightarrow |+1/2\rangle$ pair, whose oscillation frequency is $\Omega_{-1/2,+1/2} = \delta + 2D_{gs}$: within the hard-pulse model, the corresponding coefficient collapses algebraically to $X_{-1/2,+1/2} = 0$, whereas the full numerical propagation used in the main text [Figs. 2 and 4(a)] shows that this component can be one of the strongest. We therefore attribute this discrepancy to the limitations of the hard-pulse approximation in our regime, where $\omega_1$ is not asymptotically larger than the relevant splittings ($D_{gs}$ and $\delta$), and stress that Eqs. (B5)–(B13) should be interpreted as a constructive demonstration of the six-frequency cosine decomposition in Eq. (B12), not as a quantitatively accurate prediction of all branch weights under experimental conditions.

## APPENDIX C: Conservative lower-bound estimate of $T_2^*$ from positive-detuning Ramsey traces

The Ramsey trace shown in Fig. 1(e) was chosen to highlight multicomponent beating and is therefore not suitable for extracting a unique $T_2^*$. To obtain a conservative lower-bound estimate, we analyze additional positive-detuning Ramsey traces measured at $\delta/2\pi = +2$ and $+3$ MHz taken from the same dataset used in Fig. 4(b). As can already be seen from the corresponding Fourier spectra in Fig. 4(b), and explicitly for the +2 MHz case in Fig. 4(d), these positive-detuning traces are dominated by a single main Fourier component with only weak side branches. They are therefore more suitable for obtaining a conservative lower-bound estimate of $T_2^*$.

Both traces in Fig. 5 are well described over the measured 30 μs window by a single-frequency sinusoid with a weak exponential envelope. The fitted decay constants are much longer than 30 μs.

However, the oscillation amplitude does not decrease to $1/e$ within this time range. Therefore, although the fit returns a decay constant much longer than 30 μs, that value remains sensitive to the fitting model and fitting range and is therefore not taken here as a precise standalone measurement of $T_2^*$. We instead conservatively report only $T_2^* > 30$ μs.

This lower bound is sufficient to interpret the linewidth data in the main text. A dephasing time longer than 30 μs implies an inhomogeneous spin linewidth far below the MHz scale, so the cw-ODMR linewidth of $1.98 \pm 0.81$ MHz in Fig. 1(c) is clearly not Fourier limited. Rather, it reflects the particular measurement conditions used for the cw-ODMR experiment.

The present data do not identify a unique limiting mechanism of $T_2^*$. Given the isotopically engineered, unprocessed bulk sample used here, the remaining dephasing is plausibly due to weak residual nuclear-spin fluctuations together with slow local magnetic- and charge-environment noise.

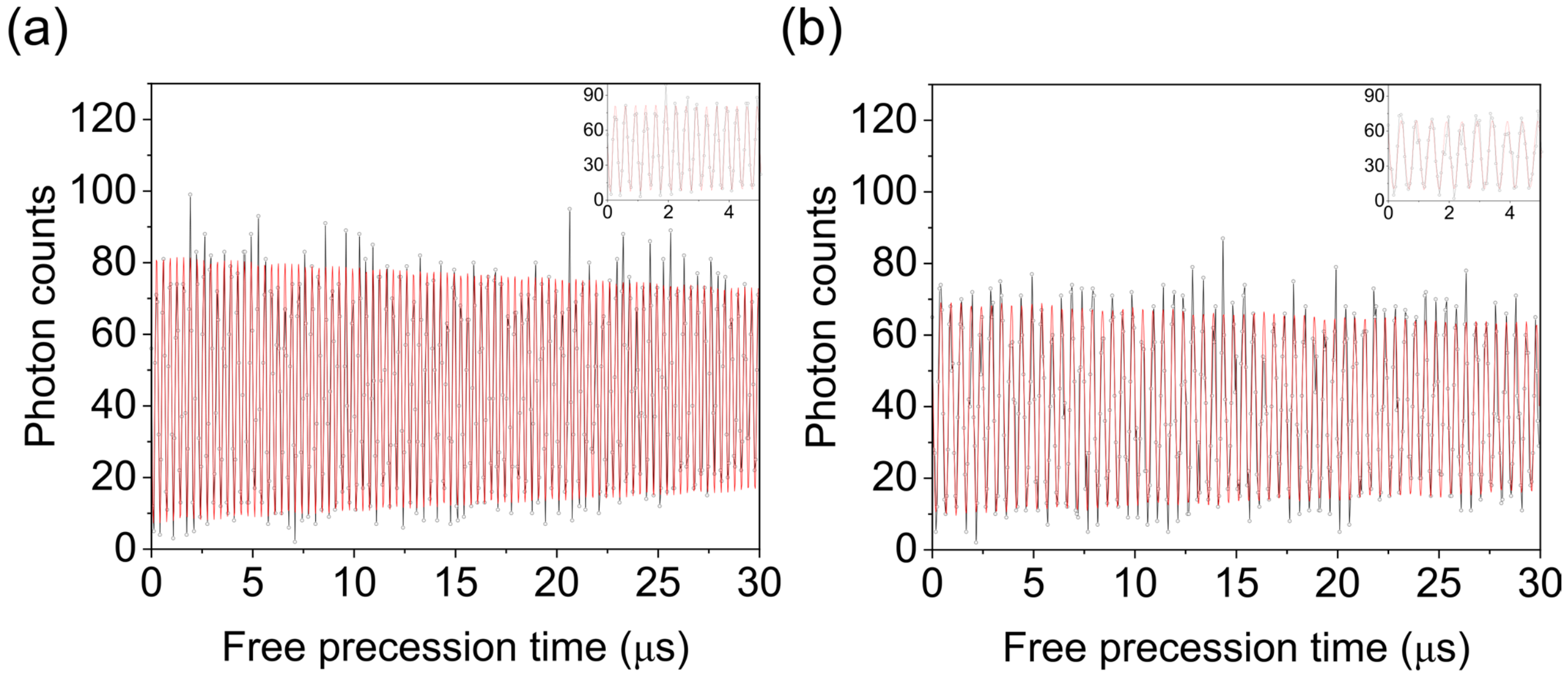

FIG. 5. Time-domain Ramsey traces at positive detunings of (a) +2 and (b) +3 MHz, together with fits to a single-frequency sinusoid multiplied by an exponential envelope. Insets show the same data on a shorter time scale. These traces are used to obtain a conservative lower bound on $T_2^*$.

## APPENDIX D: Drive-strength (bandwidth) dependence of crosstalk in Ramsey measurements

As discussed in Appendix B, the analytic model captures the frequency components of the Ramsey signal, but the relative amplitudes of the spectator-transition branches depend sensitively on the control conditions and are not accurately described within the hard-pulse approximation. To clarify the microwave-power dependence discussed in the main text, we perform additional numerical Ramsey simulations at a fixed detuning of $\delta/2\pi = -1$ MHz while varying the nominal $\pi/2$-pulse duration for the addressed transition. In the experiment, changing the microwave power changes the Rabi frequency and therefore the $\pi/2$-pulse duration required for the addressed rotation, so this provides a direct numerical representation of the experimentally relevant drive-strength dependence.

Figure 6 compares the corresponding Fourier spectra for nominal $\pi/2$-pulse durations of 25, 80, 190, 300, and 500 ns, together with the experimental trace at the same detuning extracted from Fig. 4(b). Because the spectra are shown on the same vertical scale without trace-by-trace normalization, the reduction of the spectator-induced side branches with increasing pulse duration can be compared directly.

As the pulse duration increases, the weaker spectator-transition branches are progressively suppressed, while the dominant branch associated with the addressed transition remains clearly visible. The experimentally used 80 ns pulse lies in an intermediate regime in which the additional branches remain visible, and the corresponding experimental trace is consistent with this regime. This behavior reflects the reduction of the effective control bandwidth for longer pulses, which suppresses off-resonant excitation of nonaddressed transitions. In this sense, soft-pulse control provides the most direct route to reducing crosstalk in the present Ramsey response. A broader exploration of the control landscape, together with experimental validation, is left for future work.

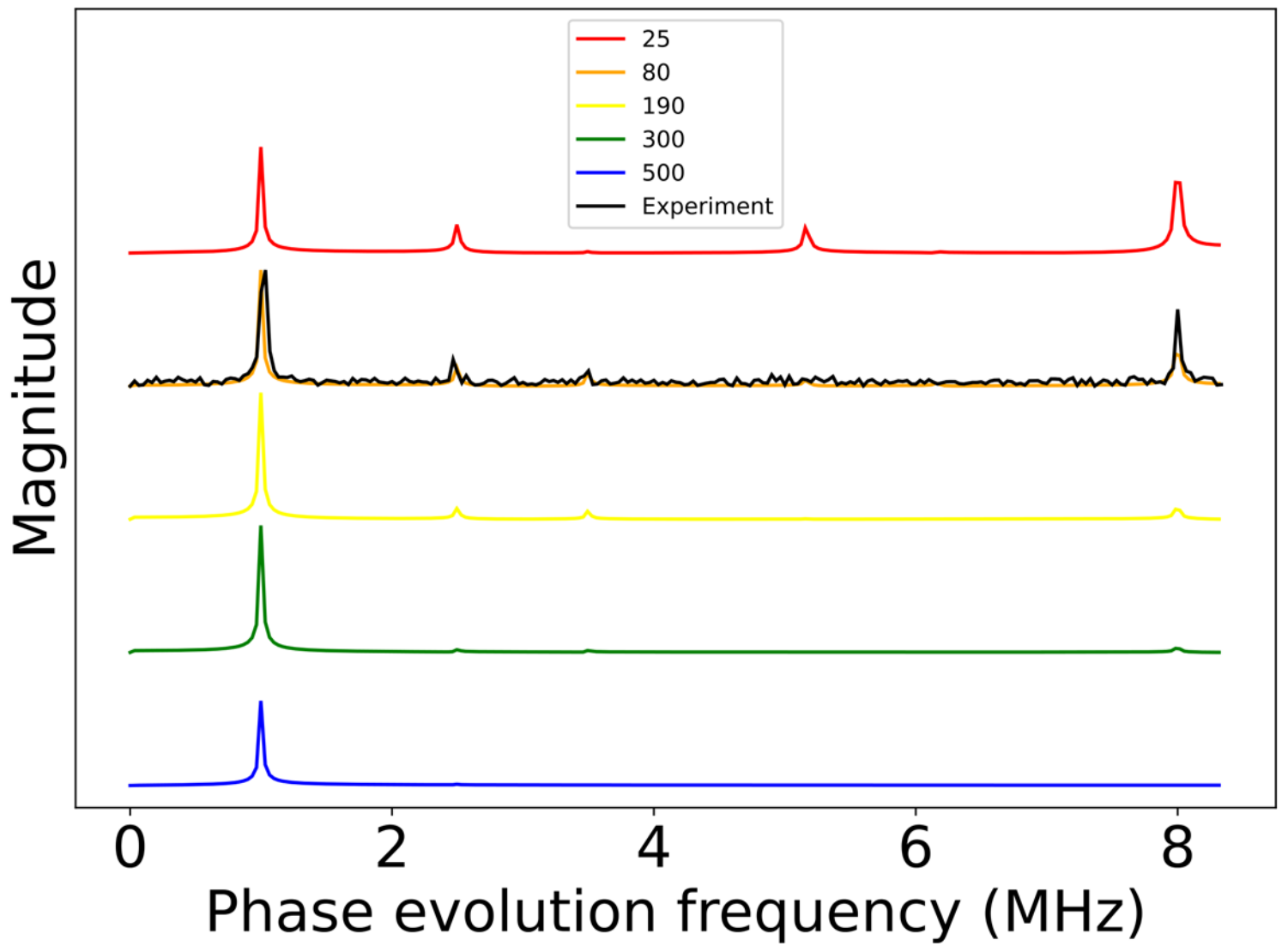

FIG. 6. Simulated Fourier spectra at $\delta/2\pi = -1$ MHz for different nominal $\pi/2$-pulse durations of the addressed transition (colored curves), together with the experimental spectrum at the same detuning extracted from Fig. 4(b) (black curve). All simulated spectra are calculated using the same experimental conditions except for the control-pulse duration, and no trace-by-trace normalization is applied. The traces are vertically offset for clarity.